\documentclass{PoS}

\title{Pion electromagnetic form factor from full lattice QCD}

\ShortTitle{Pion electromagnetic form factor from full lattice QCD}

\author{\speaker{Jonna Koponen}\\
        SUPA, School of Physics and Astronomy, University of Glasgow, Glasgow, G12 8QQ, UK\\
        E-mail: \email{jonna.koponen@glasgow.ac.uk}}

\author{Francis Bursa\\
SUPA, School of Physics and Astronomy, University of Glasgow, Glasgow, G12 8QQ, UK}

\author{Christine T. H. Davies\\
SUPA, School of Physics and Astronomy, University of Glasgow, Glasgow, G12 8QQ, UK}

\author{Rachel J. Dowdall\\
DAMTP, University of Cambridge, Wilberforce Road, Cambridge, CB3 0WA, UK}

\author{G. Peter Lepage\\
Laboratory of Elementary-Particle Physics, Cornell University, Ithaca, New York 14853, USA}

\abstract{
We present the first calculation of the pion electromagnetic 
form factor at physical light quark masses. This form factor 
parameterises the deviations from the behaviour of a point-like 
particle when a photon hits the pion. These deviations result 
from the internal structure of the pion and can thus be 
calculated in QCD. We use three sets (different lattice spacings) 
of $n_f = 2+1+1$ lattice configurations generated by the MILC 
collaboration. The Highly Improved Staggered Quark formalism 
(HISQ) is used for all of the sea and valence quarks. Using 
lattice configurations with $u$/$d$ quark masses very close to 
the physical value is a big advantage, as we avoid the chiral extrapolation. We study the shape of the vector ($f_+$) form 
factor in the $q^2$ range from $0$ to $-0.15$~GeV$^2$ and 
extract the mean square radius, $\langle r^2_v\rangle$. The 
shape of the vector form factor and the resulting radius is 
compared with experiment. We also discuss the scalar form factor 
and radius extracted from that, which is not directly accessible 
to experiment. We have also calculated the contributions from 
the disconnected diagrams to the scalar form factor at small 
$q^2$ and discuss their impact on the scalar radius 
$\langle r^2_s\rangle$.
}

\FullConference{The 33rd International Symposium on Lattice Field Theory\\
		 14 -18 July  2015\\
		 Kobe International Conference Center, Kobe, Japan}

\usepackage{amsmath}

\begin{document}

\section{Introduction}

When a photon hits a $\pi$ meson it sees the internal structure 
of the $\pi$ i.e. its quark constituents and their strong 
interaction. The electromagnetic form factor of the pion
parameterises these deviations from the behaviour of a 
point-like particle. The form factor can be calculated in 
QCD but a fully nonperturbative treatment (e.g. lattice QCD) 
is necessary. The NA7 Collaboration have determined the 
vector form factor directly experimentally from $\pi-e$
scattering~\cite{amendolia} in a model-independent way. 
Here we concentrate on calculating the form factor 
at small (negative) values of 4-momentum transfer, $q^2$,
as the experimental error is 1-1.5\% in the region up to 
$|q^2| = 0.1$~GeV$^2$ and so a lattice QCD 
calculation of the form factor there can provide a stringent 
test of QCD.

In the nonrelativistic limit, where $q^2 \approx -(\vec{q})^2$,
the electromagnetic (i.e. vector) form factor, $f_+(q^2)$, can be
viewed as the Fourier transform of the electric charge distribution. 
The mean squared radius obtained by integrating over this 
distribution is then given by 
\begin{equation}
\label{eq:rdef}
\langle r^2 \rangle  = 6 \frac{df_+(q^2)}{dq^2} \bigg|_{q^2=0}.
\end{equation}
This is adopted more generally as a definition of 
$\langle r^2 \rangle$, since it is useful to have a single number 
with which to characterise the shape of the form factor. We will 
use it here to compare different lattice calculations and also to 
compare lattice results against experiment.

\section{Lattice calculation}

Calculating the $\langle r^2 \rangle$ in lattice QCD is challenging 
as the result is very sensitive to the mass of the $\pi$. 
It has been numerically too expensive until
recently to include $u/d$ quarks with their physically very light 
masses in lattice QCD calculations: The lightest $\pi$ meson mass 
used in earlier calculations of the electromagnetic form factor has 
been in the range of 250-400 MeV, and results have had to be 
extrapolated to the physical point using chiral perturbation theory. 

Here we give results for both vector and scalar form factors for 
$\pi$ mesons made of physical $u/d$ quarks and including a fully
realistic quark content in the sea, with physical $u$, $d$, $s$ 
and $c$ quarks. We use lattice ensembles provided by the MILC
Collaboration \cite{Bazavov:2010ru, Bazavov:2012xda}, which 
allows us also to work with three different values of the 
lattice spacing --- the lattice parameters are listed in 
Table~\ref{tab:params}. This enables good control 
of systematic errors both from $m_{\pi}$ and from discretisation. 
The ensembles we use have large volumes with a minimum spatial 
size of 4.8 fm. 

\begin{table}
\centering
\label{tab:params}
\begin{tabular}{lllllllll}
\hline
Set & $a$/fm &  $am_{\ell,sea}$ & $am_{s,sea}$ & $am_{c,sea}$ & $L_s\times L_t$ & $n_{\mathrm{cfg}}$ & $T$ \\
\hline
1 & 0.1509 &  0.00235  & 0.0647  & 0.831 & 32$\times$48 & 1000 & 9,12,15 \\
\hline
2 & 0.1212 &  0.00184 & 0.0507 & 0.628 & 48$\times$64 & 1000 & 12,15,18 \\
\hline
3 & 0.0879 &  0.0012 & 0.0363 & 0.432 & 64$\times$96 & 223 & 16,21,26 \\
\hline
\end{tabular}
\caption{
The MILC gluon field ensembles used 
here~\cite{Bazavov:2010ru, Bazavov:2012xda}. Set 1 will be referred to 
``very coarse'', 2 as ``coarse'' and 3 as ``fine''. The second column 
gives the lattice spacing (determined in~\cite{Borsanyi:2012zs}). 
Columns 3, 4 and 5 give the sea quark masses
($m_u=m_d=m_{\ell}$). $L_s$ and $L_t$ are the lengths in lattice 
units in space and time directions for each lattice. The number of 
configurations that we have used in each set is given in the seventh 
column. The 8th column gives the values of the end-point of the 
3-point function, $T$, in lattice units.  
}
\end{table}

\begin{figure}
\centering
\includegraphics[width=0.9\textwidth]{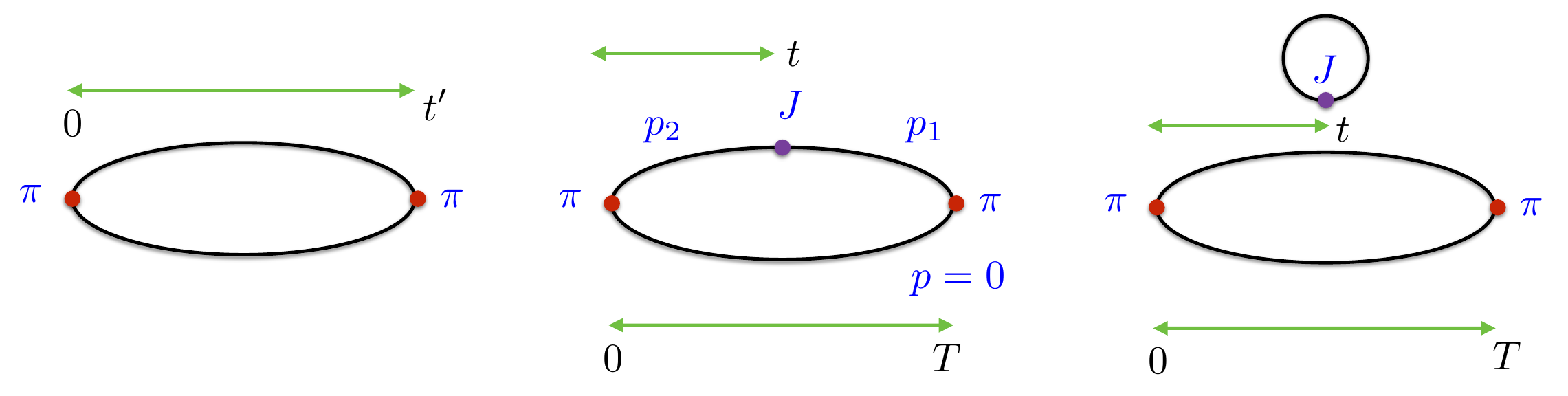}
\caption{2-point (left) and 3-point quark-line-connected (middle) 
and quark-line-disconnected (right) correlators.}
\label{fig:2pt3ptdiagram}
\end{figure}

We use the same action (HISQ) and same light quark masses for the
valence quarks as for the sea quarks, and combine the light quark
propagators into 2-point and 3-point correlators as illustrated in
Fig.~\ref{fig:2pt3ptdiagram}. 
We consider two currents $J$, a vector current and a scalar current. 
When $J$ is a vector current we need to consider only 
one 3-point diagram, the quark-line connected diagram
shown as the central diagram of Fig.~\ref{fig:2pt3ptdiagram}.
The quark-line disconnected diagram (on the right in Fig.~\ref{fig:2pt3ptdiagram}.)
vanishes for vector 
current in the ensemble average because it is odd under 
charge-conjugation, but for scalar $J$
it needs to be included.
We use twisted boundary conditions~\cite{twist} to give one
or two of the quarks spatial momentum, which allows us to
obtain the form factors at different values of $q^2$. More 
details of the calculation can be found in~\cite{pionff}.

\section{Electromagnetic form factor}
\label{sec:vec}

To extract the form factor we fit all sets of 2-point and 3-point 
correlators (at various values of spatial momenta $p_1$ and $p_2$) 
on a given ensemble simultaneously. We use a
multi-exponential fit and Bayesian methods that allow us to 
include the effect of excited states~\cite{pionff}.

We can then relate the ground state amplitude of the 3-point
correlator from the fit, $J_{0,0}$, to the matrix element between the
ground state mesons and further to the electromagnetic form factor by
\begin{equation}
f_+(q^2)(p_1+p_2)_i = 
\langle \pi(p_1) |V_i| \pi(p_2) \rangle = Z\sqrt{4E_{0}(p_1)E_{0}(p_2)}J_{0,0}(p_1,p_2),
\label{eq:me}
\end{equation}
where $E_{0}(p)$ is the ground state energy of the meson with 
momentum $p$.
We determine $Z$, the renormalisation factor, by using the fact that $f_+(0)=1$ 
for a conserved current.

\begin{figure}
\includegraphics[width=0.533\textwidth]{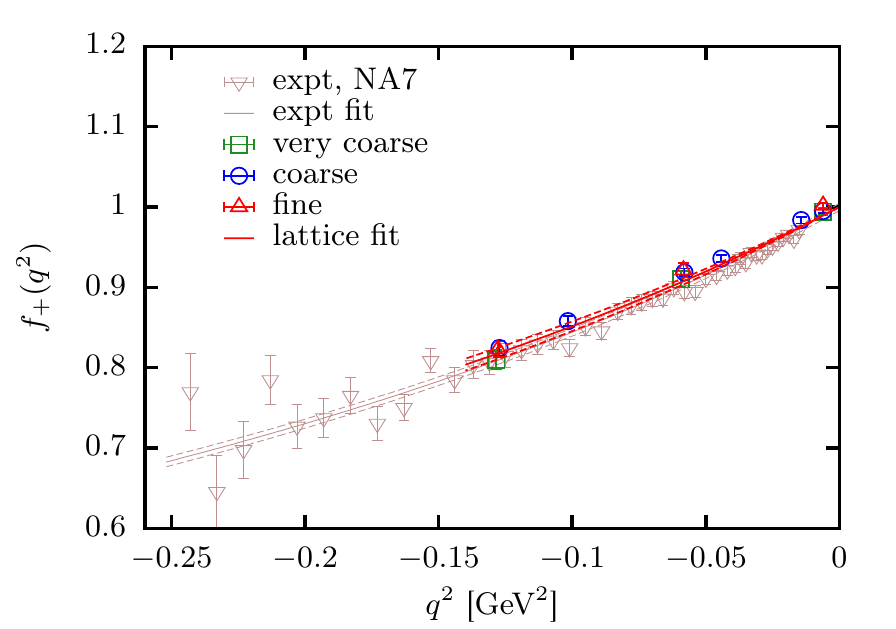}
\includegraphics[width=0.461\textwidth]{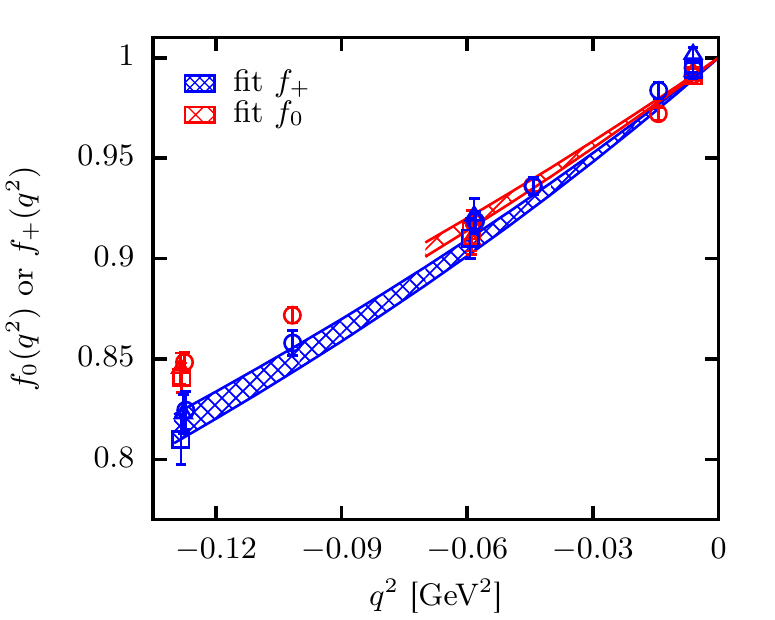}
\caption{On the left:
Lattice QCD results for the vector form factor 
on each ensemble compared directly to the  
experimental results from~\cite{amendolia}. 
Fit curves for both experiment and lattice QCD results 
are given to a `monopole' form. 
On the right:
Comparison of our lattice QCD results for the pion 
vector form factor (blue) with the connected part of the pion scalar
form factor (red). Results from set 1 are shown as open
squares, set 2 as circles and set 3 as triangles. The hashed curves give 
the fit to the form factors described in the text.
}
\label{fig:fvsexp}
\end{figure}

Our results for the electromagnetic form factor $f_+$ are plotted 
against $q^2$ in Fig.~\ref{fig:fvsexp} for all three sets along 
with the results from experiment~\cite{amendolia}. The agreement 
is very good, reflecting the fact that our results
correspond to physical $\pi$ masses.

In fitting a functional form in $q^2$ to our results to 
extract a mean squared radius, we use the same monopole form as that 
used for the experimental results~\cite{amendolia}, 
but including allowance for finite lattice spacing effects
and slight mistuning of sea quark masses~\cite{pionff}. 
We also include a logarithmic term in $m^2_{\pi}$
to make small adjustments for the fact that our u/d quark masses
are not exactly at their physical values (in fact they are slightly too low).
We plot the result of the fit in 
Fig.~\ref{fig:fvsexp} alongside with our raw data and experimental
results. Our final result for the mean square charge radius is 
$\langle r^2\rangle_V^{(\pi)}=0.403(18)(6)$~fm$^2$, where the first
error is from fitting/statistics and the second error includes all
systematic uncertainties.
More details of the fit and a full error 
budget can be found in~\cite{pionff}.

\section{Scalar form factor}


The calculation for the connected scalar form factor proceeds 
in an identical way to that of the vector form factor. 
The ground-state matrix element for the scalar current is 
related to the parameter $J_{0,0}$ extracted from 
our fits as in eq.~(\ref{eq:me}). In turn the 
matrix element is related to the form factor that 
we wish to extract by  
\begin{equation}
\label{eq:sff}
\langle \pi(p_1) |S| \pi(p_2) \rangle^{\mathrm{conn}} = Af_0^{\mathrm{conn}}(q^2),
\end{equation} 
where $A$ is a normalisation factor. 
If we had included disconnected diagrams we would 
be able to write, from the Feynman-Hellmann theorem, 
\begin{equation}
\label{eq:sff-fh}
\langle \pi(p_1) |S| \pi(p_2) \rangle = f_0(q^2)\frac{\partial M_{\pi}^2}{2\partial m_{\ell}} , 
\end{equation} 
with $f_0(0)=1$ (our scalar current is absolutely 
normalised). 
Since here we are chiefly concerned with the shape of the 
form factor, we simply treat the scalar current as requiring a 
normalisation factor, $Z_S$, and determine this from 
the requirement that also $f_0(0)=1$.  

To determine the mean squared radius associated with the 
connected scalar form factor we take the same fit as for 
the vector case, 
except that the coefficient of the chiral logarithm is 
now larger. Our final result
is 
$\langle r^2 \rangle_{S,\mathrm{conn}}^{(\pi)} = 0.349(18)(36)$~fm$^2$.
This radius has a central value that is only slightly smaller 
than the vector form factor radius as
illustrated in Fig.~\ref{fig:fvsexp}.


For the full scalar form factor we need to include the 
quark-line disconnected contribution illustrated in
Fig.~\ref{fig:2pt3ptdiagram}. 
We can then define flavour-singlet and flavour-octet 
scalar currents:
\begin{equation}
\label{eq:sdef}
S_{\mathrm{singlet}} = 2 \overline{\ell}\ell + \overline{s}s  \quad\textrm{and}\quad
S_{\mathrm{octet}} = 2 \overline{\ell}\ell - 2\overline{s}s . 
\end{equation}
Scalar form factors for these two currents are then determined 
by combining the connected scalar form factor with disconnected 
contributions in appropriate combinations from 
quark loops made from $\ell$ quarks or $s$ quarks. 

For the $q^2=0$ case it is relatively simple to calculate 
the disconnected contributions: for the $\overline{s}s$ 
scalar current this is the $\pi$ meson 
equivalent of the `strangeness in the nucleon' calculation.
The disconnected contribution for current $\overline{q}q$ is 
\begin{equation}
\label{eq:disc}
\langle \pi | S_{\overline{q}q} | \pi \rangle^{\mathrm{disc}} = \langle \pi(p) | \overline{q} q | \pi(p) \rangle - \langle \pi(p) | \pi(p) \rangle \langle \overline{q}q \rangle .
\end{equation}
This is determined from the correlation of a pion 2-point
function with source at time 0 and sink at time $T$ with a scalar 
current (condensate) summed over the timeslice at $t$. 
The second term subtracts the 
product of the vacuum expectation values of the $\pi$ 
meson correlator and the condensate.

\begin{figure}
\includegraphics[width=0.49\textwidth]{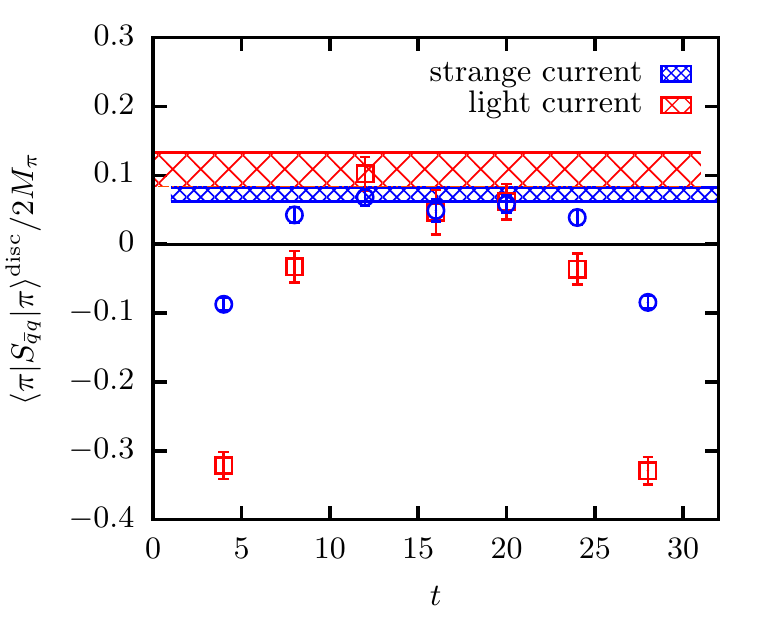}
\includegraphics[width=0.49\textwidth]{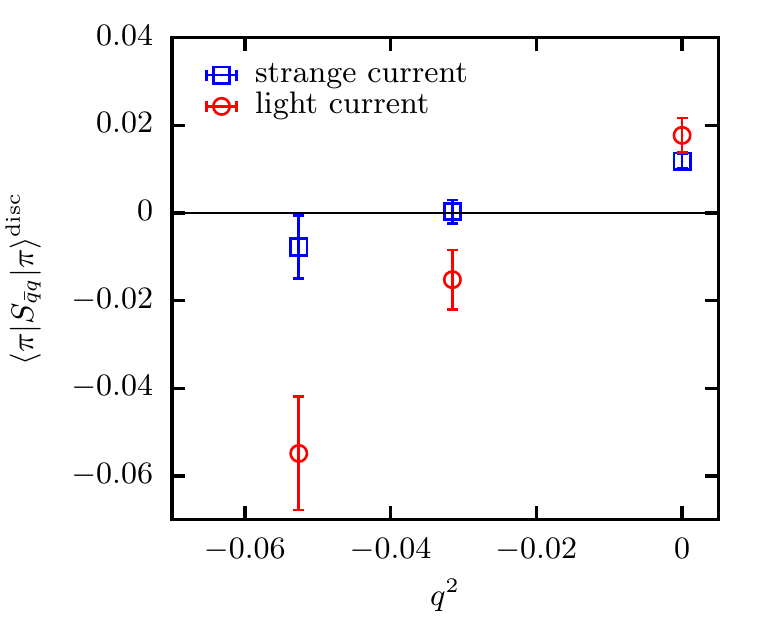}
\caption{On the left: The ratio of 3-point correlator to 2-point correlator
for the disconnected contribution for the $\ell\overline{\ell}$ 
(red circles) and $s\bar{s}$ currents (blue squares) to the scalar
form factor of the $\pi$ at $q^2=0$ on coarse lattices, set 2. 
The points are the lattice QCD results with statistical errors and the 
red and blue hashed bands show the ground-state fit result for
the $\ell\overline{\ell}$ and $s\bar{s}$ contributions, respectively.
On the right:
The $s\overline{s}$ and $\ell\overline{\ell}$ disconnected 
contributions to the scalar form factor as a function of $q^2$ for 
coarse lattices, set 2. 
}
\label{fig:f0disc_ll}
\end{figure}

The quantities required to calculate the disconnected 
contribution for the $s\overline{s}$ current 
to the scalar form factor at $q^2=0$ are then simply 
the $\pi$ meson and $\eta_s$ meson (the pseudoscalar 
$s\overline{s}$ meson) correlators. 
The $\overline{s}s$ current loop at time $t$ is obtained 
by summing over $\eta_s$ correlators with time-source $t$. 
The 3-point function that 
yields $\langle \pi | S_{\overline{s}s} | \pi \rangle$ 
of eq.~(\ref{eq:disc}) at $q^2=0$ is thus given by 
\begin{equation}
\label{eq:disccorr}
C_{3pt}(p,p,0,t,T) = - \Big\langle C_{\pi}(p,0,T)am_s\sum_{t^{\prime}}C_{\eta_s}(p,t,t^{\prime}) \Big\rangle 
+ \Big\langle C_{\pi}(p,0,T)\Big\rangle
\Big\langle am_s\sum_{t^{\prime}}C_{\eta_s}(p,t,t^{\prime}) \Big\rangle
\end{equation}
where $p=0$ and the average is over gluon configurations.
For the scalar current made of light quarks an equivalent expression holds, 
using two $\pi$ meson correlators with offset time-sources. The contribution
from the disconnected diagram is small, at the 1\% level,
compared to the connected diagram at $q^2=0$. Here we work only on coarse set 2. 

To obtain results for the disconnected contribution to the scalar 
form factor at non-zero values 
of $q^2$ we need to project onto non-zero lattice spatial 
momenta, $2\pi/L_s(n_x,n_y,n_z)$, at $T$ and $t$ in the correlators 
used in eq.~(\ref{eq:disccorr}). The statistical errors grow as 
spatial momentum is introduced so we restrict ourselves to the 
smallest non-zero lattice momenta with $(n_x,n_y,n_z)$ equal 
to $(1,0,0)$ and $(1,1,0)$ and permutations thereof.
Our results  for the disconnected diagram for both light and strange
currents at $q^2=0$ and as a function of $q^2$ are
plotted in Fig.~\ref{fig:f0disc_ll}.

To obtain the mean-square radius for the singlet and octet scalar form 
factors we must combine the connected and disconnected contributions.
We simply add the contribution from the disconnected pieces on
top of the connected results using the smallest non-zero $|q^2|$:
\begin{equation}
  \frac{|q^2|}{6}\langle r^2\rangle=\frac{|q^2|}{6}\langle r^2\rangle^{\textrm{conn}}
  \Bigg(1+\frac{f^{\textrm{disc}}_0(0)}{f^{\textrm{conn}}_0(0)}\Bigg)^{-1}
  +\frac{f^{\textrm{disc}}_0(0)-f^{\textrm{disc}}_0(q^2)}{f^{\textrm{conn}}_0(0)}
  \Bigg(1+\frac{f^{\textrm{disc}}_0(0)}{f^{\textrm{conn}}_0(0)}\Bigg)^{-1}.
\end{equation}
We find
$\langle r^2 \rangle_{S,\mathrm{singlet}}^{(\pi)} = 0.506(38)(53)$~fm$^2$
and 
$\langle r^2 \rangle_{S,\mathrm{octet}}^{(\pi)} = 0.431(38)(46)$~fm$^2$.
Here the first error is statistical and and the second error
is systematic --- see~\cite{pionff} for the full error budget.



\section{Discussion}

\begin{figure}
\includegraphics[width=0.49\textwidth]{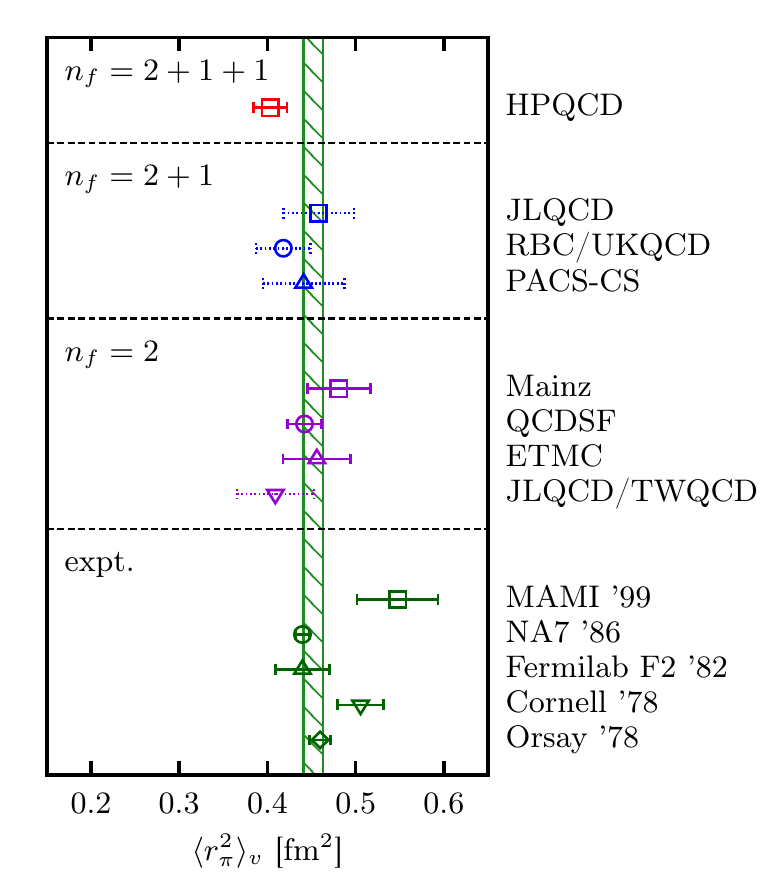}
\includegraphics[width=0.49\textwidth]{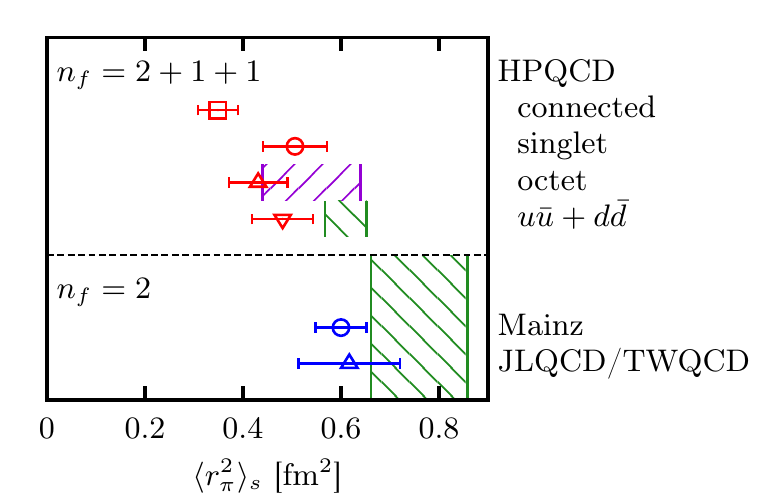}
\caption{On the left: A summary of lattice QCD results for the mean square electric 
charge radius. 
The top result is the one presented in this paper; the
$n_f=2+1$ results are from~\cite{rbcukqcd, Nguyen:2011ek, JLQCD2015}; 
and the $n_f=2$ results 
are from~\cite{Brandt:2013dua, qcdsf, etmpiff, Aoki:2009qn}.
Results that include only one value of the lattice spacing have dotted error bars.
Experimental results are from~\cite{Liesenfeld:1999mv, amendolia, Quenzer:1978qt, dally:1982zk, Bebek:1977pe}. 
The hashed vertical line gives the average from 
the Particle Data Group~\cite{PDG2014}.
On the right: A summary of lattice QCD results for the mean square scalar
radius. The HPQCD Collaboration's  results are from
this paper: ``connected'' shows the mean square radius from the quark-line connected
calculation only; ``singlet'' and ``octet'' are full calculations including
quark-line disconnected diagrams arranged in flavour-singlet or flavour-octet
currents (Eq.~\ref{eq:sdef}); for comparison, $u\bar{u}+d\bar{d}$ includes only
$u/d$ quarks in the scalar current.
The results including only $u$ and $d$ quarks in the sea ($n_f=2$) 
are from~\cite{Gulpers:2015bba,Aoki:2009qn}.
The hashed green vertical bands give the result expected from chiral perturbation
theory for $F_{\pi}/F$~\cite{FpichiPT} for $n_f = 2$ and $n_f = 2+1$
(for comparison with our $n_f = 2 + 1 + 1$ results).
The phenomenological result from $\pi-\pi$
scattering~\cite{gasserl} is very similar to the $n_f = 2+1$ green band.
The hashed purple band gives the chiral perturbation theory expectation
for the scalar octet case~\cite{FpiFKGasser}.
}
\label{fig:r2_sum}
\end{figure}

Figure~\ref{fig:r2_sum} compares results presented here for the
mean square of the pion charge radius to other lattice QCD calculations
and to experimental results. There is also a recent calculation by
B. Owen \textit{et. al.}~\cite{Owen:2015gva}, but as there is no chiral
or continuum extrapolation their results are not included in the figure.
Lattice QCD results for the mean square of the pion scalar radius
are summmarized in Fig.~\ref{fig:r2_sum} on the right.
The scalar form factor is not directly accessible in experiment so
we compare the lattice results to a phenomenological result
from $\pi-\pi$ scattering~\cite{gasserl} and to chiral perturbation theory results
for $F_{\pi}/F$~\cite{FpichiPT} and $F_K/F_{\pi}$~\cite{FpiFKGasser}. Note that
adding the disconnected diagram changes the scalar radius significantly
even though the contribution to the form factor is small.

\vspace*{3.2mm}
\noindent
\textit{Acknowledgements.} We are grateful to MILC for the use of their gauge
configurations and code.
This work was funded by STFC, NSF, the Royal Society and the Wolfson Foundation.
We used the Darwin Supercomputer of the University
of Cambridge HPC Service as part of STFC's DiRAC
facility. We are grateful to the Darwin support staff for assistance.

\end{document}